\documentclass[conference]{IEEEtran}
\IEEEoverridecommandlockouts

\usepackage{cite}
\usepackage{amsmath,amssymb,amsfonts}
\usepackage{algorithm}
\usepackage{algorithmicx}
\usepackage{graphicx}
\usepackage{textcomp}
\usepackage{xcolor}
\def\BibTeX{{\rm B\kern-.05em{\sc i\kern-.025em b}\kern-.08em
    T\kern-.1667em\lower.7ex\hbox{E}\kern-.125emX}}

\usepackage{caption}
\usepackage{url}
\usepackage{subcaption}
\usepackage{booktabs}
\usepackage{tabularx}
\usepackage{multirow}
\usepackage{multicol}
\usepackage{enumitem}
\usepackage{mathtools}

\usepackage{ragged2e}
\usepackage[all]{nowidow}
\usepackage{stfloats}
\usepackage{anyfontsize}
\usepackage{caption}
\usepackage{lipsum}
\usepackage{mdframed}
\usepackage[noend]{algpseudocode}
\usepackage{balance}
\usepackage{color}
\usepackage{colortbl}
\usepackage{mathtools}
\usepackage{colortbl}
\usepackage{listings}

\definecolor{javared}{rgb}{0.6,0,0} % for strings
\definecolor{javagreen}{rgb}{0.25,0.5,0.35} % comments
\definecolor{javapurple}{rgb}{0.5,0,0.35} % keywords
\definecolor{javadocblue}{rgb}{0.25,0.35,0.75} % javadoc
\definecolor{javagrey}{rgb}{0.46,0.45,0.48} % annotations

\newcommand\imagecaptionspace{\vspace{-0.2cm}}

\lstdefinestyle{Alg}{
  basicstyle=\ttfamily\footnotesize,
  breaklines=true,
  tabsize=2,
  mathescape,
  numbers=left,
  xleftmargin=2.5em,
  xrightmargin=0.5em,
  frame=tb,
  framexleftmargin=2em,
  emph={Algorithm,Input,Output,for,each,do,if,else,Function,while,let,be,repeat,until,return,times,and,or,break,in,then,},
  emphstyle={\textbf},
  escapechar=?,
  morecomment=[l][\color{javagreen}]{//},
  columns=flexible,
}
\usepackage{framed}
\usepackage{tikz}
\definecolor{light-gray}{gray}{0.9}
\begin{document}

\title{Learning Non-robustness  using Simulation-based Testing: a Network Traffic-shaping Case Study}

\author{
	\IEEEauthorblockN{Baharin A. Jodat, Shiva Nejati, Mehrdad Sabetzadeh}
	\IEEEauthorblockA{University of Ottawa, Ottawa, Canada\\
		Email: \{balia034, snejati,  m.sabetzadeh\}@uottawa.ca}
		\and
	\IEEEauthorblockN{Patricio Saavedra}
	\IEEEauthorblockA{RabbitRun Technologies Inc., Toronto, Canada \\
		Email: pat@rabbit.run}
}

%\author{}
%\affiliation{%
%  \institution{University of Ottawa}
%  \country{Canada}
%}
%\email{snejati@uottawa.ca}

%\author{}
%\affiliation{%
%\institution{University of Ottawa}
%  \country{Canada}
%}
%\email{}

%Baharin, Shiva, Mehrdad. 

\maketitle
\begin{abstract}
An input to a system reveals a non-robust behaviour when, by making a small change in the input, the output of the system changes from acceptable (passing) to unacceptable (failing) or vice versa. Identifying  inputs that lead to non-robust behaviours is important for many types of systems, e.g., cyber-physical and network systems, whose inputs are prone to perturbations.
In this paper, we propose an approach that combines simulation-based testing with regression tree models to generate value ranges for inputs in response to which a system is likely to exhibit non-robust behaviours. We apply our approach to a network traffic-shaping system (NTSS) -- a novel case study from the network domain. In this case study, developed and conducted in collaboration with a network solutions provider, RabbitRun Technologies, input ranges that lead to non-robustness are of interest  as a way to identify and mitigate network quality-of-service issues. We demonstrate that our approach accurately characterizes non-robust test inputs of NTSS by achieving a precision of 84\% and a recall of 100\%, significantly outperforming a standard baseline. In addition, we show that there is no statistically significant difference between the results obtained from our simulated testbed and a hardware testbed with identical configurations. Finally we describe lessons learned from our industrial collaboration, offering insights about how simulation helps discover unknown and undocumented behaviours as well as a new perspective on using non-robustness as a \hbox{measure for system re-configuration.}
\end{abstract}
%\keywords{%
%Network Traffic Shaping, Simulation, Non-robustness, Adaptive Random Testing, Regression Tree.}

%offering insights about how one can build a high-fidelity NTSS simulator to support non-robustness analysis. 

\begin{IEEEkeywords}
Simulation-based software testing, Test generation guided by machine learning, Robustness analysis, Interpretable machine learning, Network traffic shaping systems. 
\end{IEEEkeywords}

\section{Introduction}
\emph{Simulation-based testing} is concerned with developing a virtual environment that captures different components of a system including its hardware, software and network components, and using the virtual environment to test the system before it is deployed in the real-world~\cite{borg2021digital,RaquelUrtasuncompany}. Simulation-based testing  has thus far largely focused on discovering individual scenarios (tests) that can reveal system failures, e.g., system crashes or violations of some system requirement~\cite{adaptivesearch,ben2016testing,feldt2020flexible,zeller2017search,GonzalezVNBI18}. 
While revealing system failures is an essential quality assurance task, simulators can be used for a number of other important analysis tasks that are less studied in the literature. In this paper, we use simulation-based testing for characterizing a system's \emph{non-robust behaviours}. An input to a system reveals a non-robust behaviour when, by making small perturbations in the input, the output of the system changes from acceptable (passing) to unacceptable (failing) or vice versa~\cite{staliro}. Systems such as cyber-physical and network systems can be sensitive to perturbations in their input, caused by, among other factors, uncertainty in the environment, evolving system-usage patterns, internal computation errors, and network degradation. For these systems, it is important to be able to identify test inputs that elicit non-robust  behaviours.

Feldt and Yoo~\cite{feldt2020flexible} observe that in the existing literature on software testing, a large number of system executions are typically performed merely to produce a single test at the end. In the context of simulation-based testing, this means that  a large number of often compute-intensive simulations are left unused, thus wasting time and resources. Some recent research proposes to use these otherwise wasted simulation results for building machine learning models~\cite{raja2018,miningassumptions} or generative models~\cite{feldt2013finding,feldt2020flexible,jia}. These models are  built incrementally by a test generation algorithm and approximate the entire or a part of the system's input space. The main usage of these models is to guide test generation (e.g., by  exploring regions that are more likely to reveal failures), and to provide additional feedback, e.g., in the form of failure models~\cite{gopinath2020abstracting}.
%since they  specify constraints on  system inputs under which the system is likely to fail. 

In this paper, we follow the same line of research and propose an approach that combines machine learning and adaptive random testing to generate value ranges for test inputs in response to which the system is likely to exhibit non-robustness. We apply our approach to  a \emph{network traffic-shaping system (NTSS)} -- a novel case study from the network domain~\cite{cakepaper,hoiland2015good,hoeiland2018flow}. Traffic shaping is an advanced technique to improve the quality of transmission for voice, video and other types of streaming traffic. To configure an NTSS in a way that ensures high quality of network transmission while maximizing bandwidth utilization, we need to identify input ranges that make the  NTSS non-robust.

In collaboration with our industry partner, RabbitRun Technologies (\url{https://www.rabbit.run}), we develop a simulation environment to test NTSS. We then present Non-Robustn\textbf{E}ss A\textbf{N}alysis for t\textbf{R}aff\textbf{IC} S\textbf{H}aping (ENRICH), a method to approximate input ranges that likely lead to non-robust NTSS behaviours. ENRICH implements an adaptive random testing algorithm based on our NTSS simulator. The test cases generated by adaptive random testing are used to train a machine learning regression tree from which the areas in the input search space that include system's non-robust behaviour are inferred. These areas are then passed to the adaptive testing algorithm to focus  test generation in these inferred areas, since these areas likely include  inputs that make the system non-robust. The iterative test generation and regression tree model refinement continues until 
the computational budget is exhausted. The final regression tree will then be used to infer value ranges for the NTSS inputs that lead to  non-robustness. 

The regression tree model generated  by ENRICH in the first iteration is trained  on evenly distributed samples in the entire search space, thus yielding an \emph{explorative} (global) view. In contrast, the models generated in later iterations become more focused on  inputs that likely make the system non-robust. These models provide an \emph{exploitative} view on the desired regions of the search space. 
It is difficult to accurately approximate the whole search space relying on explorative views only. As shown in earlier research~\cite{adaptivesearch, wang2017committee},  the gradual move from an explorative to an exploitative view, as adopted by ENRICH, is more effective at inferring promising areas of the search space, \hbox{i.e., non-robust regions in the context of our work.}

We evaluate ENRICH on an NTSS setup recommended by RabbitRun. We compare ENRICH with a standard baseline based on random testing. The baseline infers non-robust test inputs using a regression tree model built based on samples uniformly selected from the search space (i.e., an explorative model without the gradual refinement into an exploitative model). Our results show that ENRICH significantly outperforms the baseline in generating and characterizing non-robust test inputs for NTSS. In particular, \emph{ENRICH is able to identify non-robust test inputs with a precision of $84$\% and a recall of $100$\% while yielding a significantly higher overall accuracy than the baseline. 
In addition, our results show that there is no statistically significant difference between the test results obtained from our simulator and the results obtained by executing the same tests on a physical (hardware) testbed.}

\vspace*{.5em}\emph{Contributions.} We make the following contributions:

$\bullet$ We introduce the problem of capturing non-robust test inputs for network traffic-shaping systems (Section~\ref{sec:motivate}). 
    
$\bullet$ We build an  industry-strength simulation environment for NTSS (Section~\ref{simplat}).  Detailed instructions for building the 
simulation environment are publicly available~\cite{enrichgithub}.  
    
$\bullet$ We present ENRICH -- an approach to automatically infer input ranges that likely lead to non-robust NTSS behaviours (Section~\ref{sec:approach}). 

$\bullet$ We evaluate the accuracy of ENRICH and our  NTSS simulator (Section~\ref{sec:eval}). Our experimental results are publicly available~\cite{enrichgithub}. 

$\bullet$ We reflect on the lessons learned from our collaboration with RabbitRun Technologies (Section~\ref{sec:lesson}).

\vspace*{.5em}\emph{Structure.} Section~\ref{sec:motivate} 
motivates non-robustness analysis for NTSS. Section~\ref{simplat} describes our NTSS simulator. Section~\ref{sec:approach} presents our approach for characterizing non-robust test inputs. Section~\ref{sec:eval} describes our evaluation. Section~\ref{sec:related} compares with related work.  Section~\ref{sec:lesson} outlines our lessons learned. Section~\ref{sec:con} concludes the paper.

%{\color{red} WE NEED AN EXAMPLE HERE. For example, GIVEN EXAMPLE OF Non-robust behaviour.} 

%helpful for early fault resolution, is still quite limited in terms of the amount of feedback we can possibly provide engineers with using simulators

%For example, many simulation-based testing techniques in the context of the automotive and self-driving systems employ fuzz testing and search-based testing approaches  that aim to generate \emph{individual} test inputs that reveal some system failure~\cite{??}. 

%Our research uses intermediate test executions to generate boundary behaviour.

\section{Industrial Context and Motivation}
\label{sec:motivate}
RabbitRun Technologies (RRT) provides advanced network connectivity solutions for the Small-Office/Home-Office  (SOHO) market. The SOHO market has been growing in importance in recent years, and has been crucial during the COVID-19 pandemic with a major part of the workforce needing to work from home. RRT employs a well-established technique, known as traffic shaping~\cite{cakepaper}, and develops a network traffic-shaping system (NTSS) to support high-quality connectivity for SOHO, in particular for \emph{real-time streaming} applications such as voice and video. Traffic shaping is applied at the edge of the network and on routers to control the \emph{outbound} traffic, i.e., the traffic travelling from equipment out onto the Internet.  %Briefly, provided with a specified amount of bandwidth for transmission, an NTSS  delays the flow of certain types of packets in order to ensure a higher quality of transmission for packets with high priority which are often set to be voice and video packets. As a result, traffic shaping  enables high-quality transmission of several parallel voice and video streaming applications (e.g., several concurrent Zoom calls) and is crucial for connectivity in SOHO settings.

%Specifically, NTSS prioritizes  \emph{real-time streaming} applications such as voice and video to ensure their high quality. 

%Traffic shaping prioritizes transmitting voice and video and ensures that their packets are transmitted in the same sequence that they were sent without any loss. 

%see comment

%In fact, traffic shaping becomes critical when network uplinks become overburdened with data being delivered out of an interface %https://www.techtarget.com/searchnetworking/definition/traffic-shaping.

\begin{figure}[]
    \centering
    \includegraphics[width=\columnwidth]{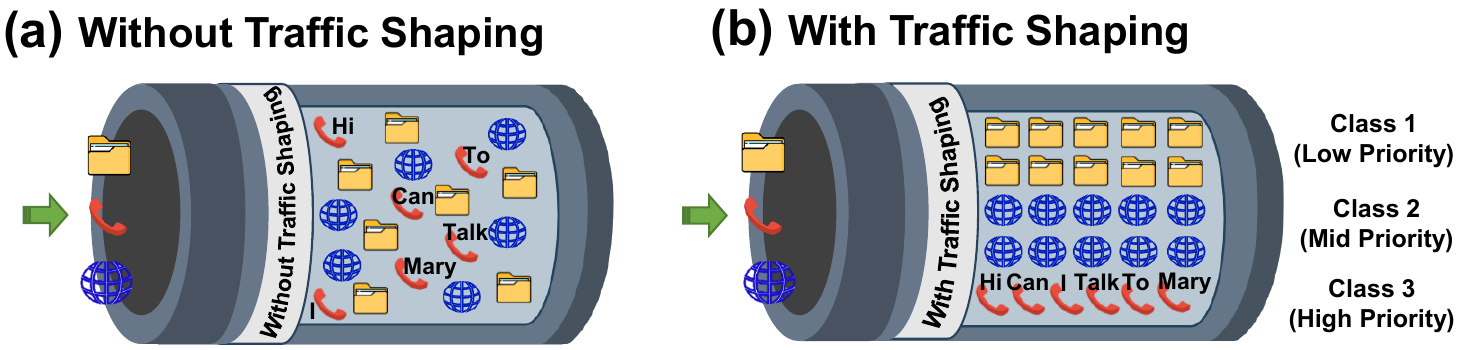}
    \vspace*{-.3cm}
    \caption{Illustration of network behaviour (a) without traffic shaping, and (b) with traffic shaping~\cite{trafficshapingfigure}.}\label{fig:trafficshaping}
    \vspace*{-.2cm}
\end{figure}

Figure~\ref{fig:trafficshaping}  illustrates the working of an NTSS. If we transmit a voice message (e.g., ``Hi, Can I talk to Mary?'') without  traffic shaping, the voice packets may be mixed with other packets. This makes it difficult to transmit the voice packets fast while keeping them in the same sequence  they were sent (Figure~\ref{fig:trafficshaping}(a)). Consequently,  the receiver of the voice message, who needs to process the voice packets as they arrive, may feel that the voice is delayed or the conversation may be cut out for some seconds. To avoid the degradation of network quality for streaming applications, one can have an NTSS manage outbound traffic on routers. As shown in Figure~\ref{fig:trafficshaping}(b), an NTSS divides the available bandwidth into a number of classes with different bandwidth thresholds and priorities. Voice and video packets are then allocated to a specific class where they can be transmitted together and in a timely manner. 
%Note that for voice and video, even a few dropped or delayed  packets  may be quickly sensed by users since receivers of voice and video packets process them instantaneously. This is different from sending an email or a file, for example, since the network has a chance to resend the lost email or file packets without the receivers noticing a delay.

%https://net2phone.ca/resources/blog/what-is-traffic-shaping-and-why-do-i-need-it

NTSS can be set up  with different numbers of classes and different rules for specifying bandwidth thresholds and priorities \cite{cakepaper, cakegithub}. The simple NTSS example in Figure~\ref{fig:trafficshaping}(b) has  three classes and is configured such that voice and video applications are allocated to \texttt{class3} -- the highest-priority class. Browsing requests are mapped to \texttt{class2} with middle-level priority, and file sharing requests are mapped to \texttt{class1} -- the lowest-priority class. For a complex NTSS setup, engineers need to have tools to help them configure the NTSS in the most optimal way. That is, they need to know which applications should be mapped to which class to ensure the highest quality of service for streaming traffic without starving any other request types (e.g., browsing and file sharing). 

%An analogy that is often made is dividing the highway traffic into a fast lane and a commuter lane. The fast lane is only one lane (low threshold), but since it is reserved for specific vehicles (e.g., emergency services and electric vehicles), it still allows their fast transmit.

%Over the years a number of complex traffic shaping algorithms and policies have been developed in the network literature~\cite{??}. At the same time, our societies are becoming even more connected and requiring the network systems to transmit even more and larger amount of data. This changes the expectations that we have from a traffic shaping algorithm. For example, a few years ago, to have good quality for voice over IP, we only needed to create a class with 10\% of the total bandwidth to transmit voice quickly. But, the needs for transmitting voice and video is now increasing which requires to configure traffic shaping with more classes and creating more sophisticated priority rules. 

\begin{figure}[t]
    \centering
    \includegraphics[width=\columnwidth]{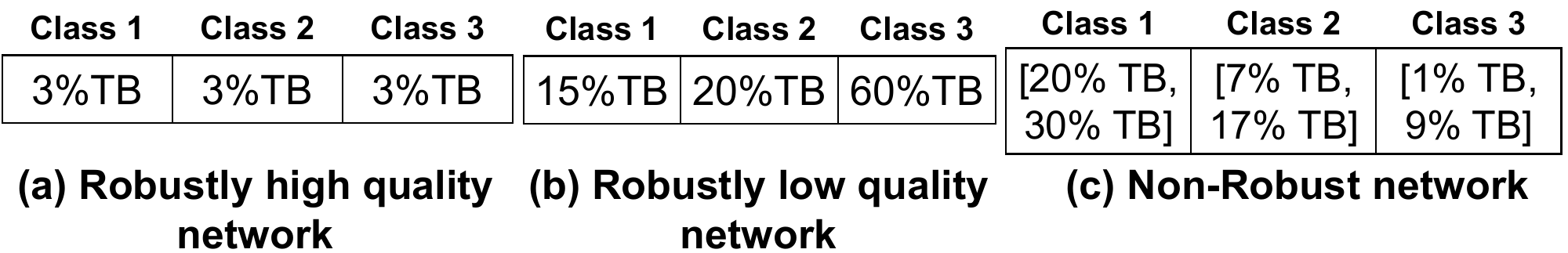}
    \captionsetup{skip=0pt}
    \setlength\belowcaptionskip{0pt}
    \caption{Input vectors for a  network traffic-shaping system (NTSS) with three classes and a total available bandwidth denoted by TB: (a)~low-bandwidth traffic (9\% of TB) leading to a robustly high-quality network; (b)~high-bandwidth traffic (95\% of TB) leading to a robustly low-quality network; and (c)~bandwidth ranges leading to a  non-robust network.}\label{fig:example}
    \vspace*{-.2cm}
    
\end{figure}

To help engineers  optimally configure an NTSS, we need the ability to approximate the boundary between robust and non-robust behaviours. Figure~\ref{fig:example} exemplifies robust versus non-robust behaviours for an NTSS with three classes whose total available bandwidth is denoted by TB. In this system, for low outbound traffic (e.g., $9$\% of TB as shown in Figure~\ref{fig:example}(a)), users experience good-quality connection, and minor changes in the input do not affect the quality. On the other hand, for high outbound traffic (e.g., $95$\% of TB as shown in Figure~\ref{fig:example}(b)), users  experience low-quality traffic that does not improve with minor changes in the input either. In other words, the examples in Figures~\ref{fig:example}(a) and (b) represent robust behaviours where for the former, the network quality is robustly good, and for the latter, the network quality is robustly bad. As one crosses between good-quality network situations to poor-quality ones, there is a sizable range of input traffic values where the system is volatile and where an input traffic stream with acceptable quality may turn bad due to small fluctuations in the input traffic ranges. 
Figures~\ref{fig:example}(c) illustrates  example input ranges that may lead to \emph{non-robust} behaviours in an NTSS. 

Knowing about input ranges for which an NTSS likely becomes non-robust can help engineers in the following ways: The ranges can guide engineers  in better mapping applications to classes. For example, if the engineers know that the non-robust range for \texttt{class1} is around 100mb/s and for \texttt{class2} around 150mb/s, they can assign
applications to \texttt{class1} (resp. \texttt{class2}) with bandwidth values below 100mb/s (resp. 150mb/s).  In this way, these ranges help solve the trade-off between optimal utilization versus providing acceptable and robust quality. Another use case for these input ranges -- a topic that we are currently exploring with our partner -- is to devise run-time adaptation mechanisms that can steer the system away from non-robust regions, e.g., by dynamically reclassifying the traffic originating from different applications.

%The rest of this paper presents our simulator for NTSS and our approach to approximating value ranges for NTSS inputs that lead to non-robustness.  

%(2)  Given the large input space of such systems, the engineers cannot possibly test the quality of their system for its entire input space. Knowing the ranges for non-robust behaviours helps engineers  focus on testing their systems within those ranges. 

%In this paper, we use adaptive random  testing for software testing combined with decision tree regression to derive ranges for non-robust behaviours of NTSS. 

%In a nutshell, traffic shaping comprises two principles: The total bandwidth is divided into a number of classes. Each class has a priority and a bandwidth threshold. The applications and packet types are divided between these classes according to a pre-defined mapping. The main idea is to provide different types of applications with different needs and characteristics different bandwidth threshold and different priority. This is mainly to  ensure that the quality of voice and video streams are maintained at all time. 

% https://www.mushroomnetworks.com/blog/bufferbloat-what-is-it-and-why-you-or-your-vendor-should-care/
 
% https://net2phone.ca/resources/blog/what-is-traffic-shaping-and-why-do-i-need-it
\section{Our NTSS Simulator}
\label{simplat}
\begin{figure}[t]
    \centering
    \includegraphics[width=\columnwidth]{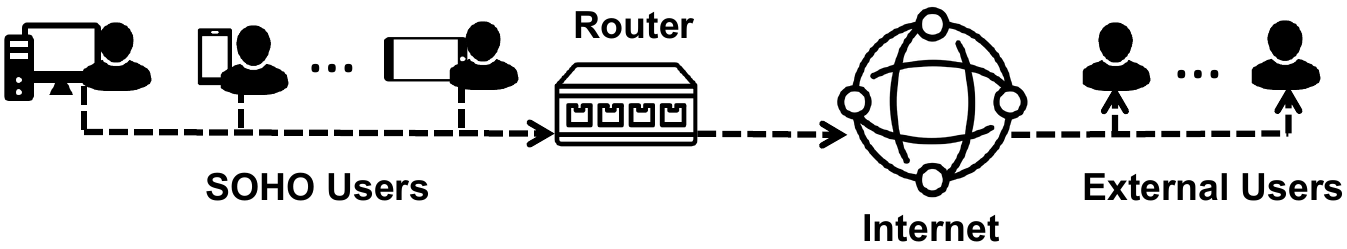}
    \caption{\protect \raggedright Overview of  Small-Office/Home-Office (SOHO). %The direction of network flows is from SOHO users to external users.
    }
    \label{fig:architecturegoals}
    \vspace{-0.3cm}
\end{figure}

We develop an NTSS simulator to serve as our test-execution tool and enable the simulation of  various network-usage scenarios in a generic SOHO setting.  Figure~\ref{fig:architecturegoals} shows the physical view of a SOHO setting where SOHO users are connected to a router via different types of devices (PCs, phones, tablets, etc.). Through an internet modem, the router sends  data packets from the SOHO users to some external users' IPs specified in the packets' headers. Note that an NTSS controls the outbound traffic only, i.e., the direction from SOHO users to the external users in Figure~\ref{fig:architecturegoals}. Hence,   our simulator is focused on generating outbound data flows.

%Figure~\ref{fig:architecturegoals} shows the architecture of a SOHO instance that we consider in our work. As shown in the figure, a number of SOHO users (three in the figure) are connected to a router via a network switch. The router then connects to other users via Internet or other networks. 
%MOVE TO SECTION 2: In fact, traffic shaping becomes critical when network uplinks become overburdened with data being delivered out of an interface %https://www.techtarget.com/searchnetworking/definition/traffic-shaping.
%"Shaping can only occur on packets that are leaving an interface as opposed to coming into the interface".https://www.techtarget.com/searchnetworking/definition/traffic-shaping
Our simulator, named \emph{SOHOSim},  generates outbound traffic streams in the SOHO setup of Figure~\ref{fig:architecturegoals} and measures the quality of the network. Using virtual machines, SOHOSim captures all the elements in Figure~\ref{fig:architecturegoals}, namely SOHO users, the router and external users  (consult~\cite{enrichgithub} for more details). SOHOSim can simulate parallel data flows and streams (e.g., Zoom calls, VOIP, or gaming) sent by several simultaneous SOHO users. This enables us to  run a large number of tests involving several parallel network flows sent by several users capturing many different realistic situations. Testing such situations on a physical setup is expensive and  time-consuming. SOHOSim can measure network quality for SOHO users to, for example,  determine if the users are  experiencing network problems  (e.g., choppy Zoom calls). In this way, the testing performed using SOHOSim allows us to determine whether an NTSS installed on the router is configured properly.

\begin{figure}
    \centering
    \includegraphics[width=.8\columnwidth]{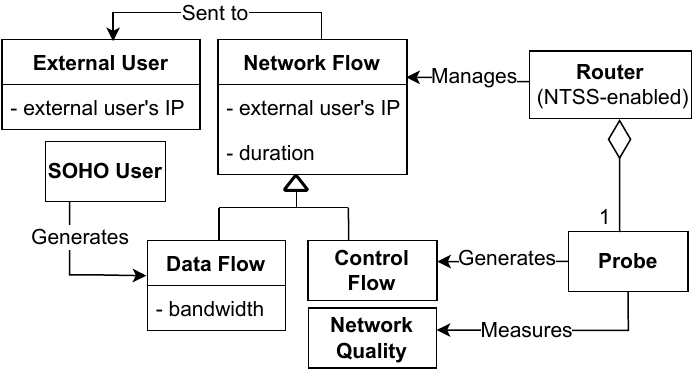}
      \vspace*{-.2cm}
    \caption{Conceptual view of our NTSS simulator, SOHOSim. %capturing the SOHO setting in Figure~\ref{fig:architecturegoals} and used for NTSS testing.
    }\label{fig:virtualsetup}
    \vspace*{-.4cm}
\end{figure}

A conceptual model for SOHOSim is shown in Figure~\ref{fig:virtualsetup}. For the purpose of testing an NTSS, our simulator needs to generate two types of flows: \emph{data flows} and \emph{control flows}. Each data flow has a specified bandwidth,  duration, and a destination IP. Each data flow is produced by a SOHO user and sent to a specific external user as specified by the destination IP of the data flow.  Control flows are used internally by SOHOSim to measure network quality.  SOHOSim has a single router entity that manages network flows, both data and control. The router is configured to use an NTSS to separate the flows into different classes and ensure high quality network connectivity as discussed in Section~\ref{sec:motivate}. The \emph{probe} entity, which operates on the router, generates control flows and measures the quality metrics. Control flows are low-bandwidth,  sent by the probe and received by the same external users that receive data flows.  In contrast to data flows, control flows are echoed back to the router so that the probe can compute quality metrics such as latency, loss and jitter. For control flows, the bandwidth is fixed and set to a low value. The duration and the destination of control flows are the same as those of data flows. 

To use SOHOSim for testing an NTSS, we generate parallel data flows from SOHO users to different external users such that all flows pass through the router. 
The NTSS installed on our (virtual) router then divides these flows into separate classes. 
Using SOHOSim, we measure the network quality for each NTSS class as we further discuss in the next section.

\section{Approach}
\label{sec:approach}
%Introduce our approach. 
In this section, we present our approach
for capturing non-robust test inputs in NTSS. Our approach 
leverages SOHOSim for test execution. Section~\ref{subsec:domain} defines  test inputs and outputs for NTSS. Section~\ref{subsec:fitness} introduces our robustness measure that enables us to distinguish between robust and non-robust behaviours. Section~\ref{subsec:generation} presents Non-Robustn\textbf{E}ss A\textbf{N}alysis for t\textbf{R}aff\textbf{IC} S\textbf{H}aping (ENRICH) -- our proposed approach for identifying non-robust test inputs in NTSS. %ENRICH utilizes our robustness measure alongside adaptive random testing~\cite{metaheuristicsbook} and regression trees~\cite{dataminingbook}.

\subsection{Test Input/Output Formalization}
\label{subsec:domain}
An NTSS consists  of a set $C = \{c_1, \ldots, c_n\}$ of $n$ classes. Each NTSS class $c_i$ has a bandwidth range $[0..\mathit{bwR}_i]$ and a  \emph{priority}. We assume that the class indices represent their priority order with $c_n$ being the highest-priority and $c_1$ being the lowest-priority class. Let $c_i$ and $c_j$ be a pair of classes such that $c_i$ has higher priority than $c_j$ (i.e., $i > j$).  A traffic-shaping algorithm  provides better network quality (e.g., lower latency and loss) for the flows that go through $c_i$ compared to those going through $c_j$ as long as the bandwidth of the flows going through $c_i$ (resp. $c_j$) remain below the maximum bandwidth of $c_i$ (resp. $c_j$)~\cite{CAKE}.

%As shown in Figure~\ref{fig:architecturegoals}, in order to test NTSS, SOHOSim generates traffics from (virtual) \textsf{SOHO users} to some traffic destination by passing the traffic through the (virtual) router running NTSS.  As discussed in Sections~\ref{sec:motivate} and \ref{??}, our characterization of non-robust behaviours enables engineers to optimize such mappings. 

Each test input for an NTSS is a tuple $(\mathit{tr}_1, \ldots, \mathit{tr}_n)$ where each $\mathit{tr}_i$ is the bandwidth of the flow going through class $c_i$.  SOHOSim is able to generate flows with different bandwidths for different NTSS classes.  The range for each $\mathit{tr_i}$ is $\mathit{[0.. \mathit{bwR}_i]}$. That is, we generate test inputs such that the flow in each class $c_i$ remains below the maximum bandwidth of $c_i$. %The bandwidth threshold of each class is determined based on NTSS algorithm policy~\cite{??}. 

A standard and well-known metric used in the network community to quantify network quality is \emph{Mean Opinion Score (MOS)}~\cite{mos}. The MOS value is a real number ranging from 1.0 to 5.0, where 1.0 indicates the lowest quality  and 5.0 indicates the best quality. 
We use MOS to measure network quality  for each NTSS class. Specifically, for each test input $(\mathit{tr}_1, \ldots, \mathit{tr}_n)$, SOHOSim measures the MOS value corresponding to each input flow 
$\mathit{tr}_i$ passing through 
class $c_i$.  For example, suppose we test a four-class NTSS using a test input $(240, 230, 200 ,100)$, and suppose SOHOSim measures MOS values $2.51$, $3.3$, $4.41$, $4.49$ for $c_1$, $c_2$, $c_3$ and $c_4$,  respectively. That is, the tuple $(2.51, 3.3, 4.41, 4.49)$ is the  output corresponding to the test input $(240, 230, 200, 100)$.

For each class $c_i$, engineers can determine a threshold for the MOS value measured for that class in order to differentiate  between \emph{good} (acceptable) network quality and \emph{bad} (unacceptable) network quality. We refer to this threshold as the \emph{MOS threshold} and denote it by $\mathit{mosTh}$. The MOS threshold for each class can be determined based on domain knowledge and the configurations of an NTSS.  

%For example, engineers may consider a higher MOS threshold for a class that is used for voice and video, and a lower MOS threshold for a class used for email and file transfer. This is because they usually want to maintain a higher quality for streaming applications (voice and video) compared to other applications (e.g., file transfer). 

\subsection{Robustness Measure}
\label{subsec:fitness}
As discussed in section~\ref{subsec:domain}, for each test input $(tr_1, \ldots, tr_n)$, SOHOSim measures a tuple $(\mathit{mos}_1, \ldots, \mathit{mos}_n)$ as the test output such that each $\mathit{mos}_i$ specifies the network  quality  for class $c_i$. In addition, for each $c_i$, we have a MOS threshold  $\mathit{mosTh}_i$. If $\mathit{mos}_i$ is higher than $\mathit{mosTh}_i$, the network quality at class $c_i$ is acceptable (good); otherwise, the quality is unacceptable (low).

To be able to identify non-robust test inputs, it is not sufficient to determine the quality for each class individually. Instead, we need an aggregated measure that can determine, for a given test input, whether or not the NTSS performance as a whole (i.e., for all the classes) is acceptable. To do so, we need to combine the $n$ MOS outputs obtained for a test input to compute a single measure. We adopt an approach that has been used in the search-based testing literature to define hybrid test objectives and combine several metrics simultaneously~\cite{fraser2012whole, mcminn2004search}. This approach allows one to not only aggregate different measures into one, but also to retain the priority of each measure in the aggregated measure. Specifically, for an NTSS, low quality of traffic on a higher-priority class is worse than low quality of traffic on a lower-priority class. We combine the MOS values measured for different NTSS classes in such a way that the aggregated measure preserves the priority of the classes. We refer to our single measure as  \emph{robustness measure}.

To define our robustness measure, we first normalize the MOS values obtained for each class. We use a well-known rational function $\omega(x) = x / (x + 1)$ for normalization~\cite{arcuri2013really}. We denote by $\overline{mos}_i$ the normalized form of each MOS value $\mathit{mos}_i$.  
For a given test output $(\mathit{mos}_1, \ldots, \mathit{mos}_n)$, we denote our robustness measure by $ \mathcal{R}(\overline{mos}_1, \ldots, \overline{mos}_n) $ and define it as follows: 

\vspace*{1em}

\hspace*{-.3cm}\scalebox{0.7}{
$
    %\mathcal{R} = 
    \begin{cases}
        \overline{mos}_n & \mbox{if } \bigwedge_{i=1..n} \overline{mos}_i < \overline{mosTh}_i, \\
        1 + \overline{mos}_{n-1} & \mbox{if } \overline{mos}_n \geq \overline{mosTh}_n  \land \bigwedge_{i=1..n-1} \overline{mos}_i < \overline{mosTh}_i , \\
        2 + \overline{mos}_{n-2} & \mbox{if } \bigwedge_{i\in\{n-1,n\}} \overline{mos}_i \geq \overline{mosTh}_i  \land   \bigwedge_{i\in \{1..n-2\}} \overline{mos}_i < \overline{mosTh}_i ,\\
        \ldots & \ldots \\
        n & \mbox{if } \bigwedge_{i=1..n} \overline{mos}_i \geq \overline{mosTh}_i
    \end{cases}
$
}
\vspace*{1em}

where $\overline{mosTh}_i$ is the normalized form of the MOS threshold $mosTh_i$. The robustness measure $\mathcal{R}$ is within the range $[0.5, n]$ since MOS values cannot go below $1.0$ (see Section~\ref{subsec:domain}); as such, 
$\overline{mos}_i$ values cannot go below $0.5$. A robustness value of $0.5$ indicates that the network quality for the highest-priority class is low; a robustness value of $n$ means that the network quality for all the classes is high. More precisely, the robustness measure is interpreted as follows: 

\vspace*{.2cm}
\scalebox{0.95}{$\begin{array}{lcl}
0.5 \leq \mathcal{R} < 1  &\Rightarrow&  c_1\ldots c_{n} \mbox{ are low quality}\\
1.5 \leq \mathcal{R} < 2 &\Rightarrow& c_1\ldots c_{n-1} \mbox{ are low quality}, \\
& & c_n \mbox{ is high quality} \\
\ldots & & \ldots\\
 n-1 +\frac{1}{2} \leq \mathcal{R} < n &\Rightarrow& c_1 \mbox{ is low quality}, \\
 & & c_2\ldots c_n \mbox{ are high quality}\\
\mathcal{R} = n &\Rightarrow& c_1\ldots c_{n} \mbox{ are high quality}
\end{array}$}
\vspace*{.2cm}

 If the robustness measure for test $i$ is higher than that for test $j$, then the network quality is higher for test $i$ than for test $j$.  To differentiate between acceptable and unacceptable behaviours of an NTSS, engineers can set a threshold on the robustness measure; we denote this by $\mathit{rbTh}$. %The threshold can be set to any value between $0.5$  and $n$ depending on the specific usage of an NTSS. For example, setting $\mathit{rbTh}$ to $n$ means that the network quality is acceptable only when  all classes are at their best quality, while $\mathit{rbTh} = 2.5$ indicates that the NTSS performance is acceptable as long as the two high-priority classes $n$ and $n-1$ are preforming well. 
 
\begin{figure}[t]
    \centering
    \includegraphics[width=0.65\columnwidth]{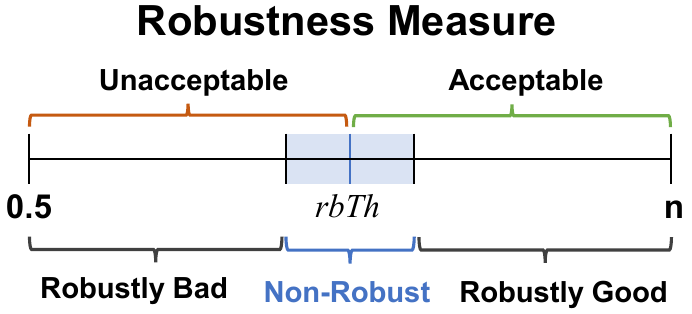}
    \imagecaptionspace
    \caption{Illustrating the relationship between  robust versus non-robust and acceptable versus unacceptable sub-ranges within the robustness measure range.}\label{fig:robustnessmeasurerange}
    \vspace{-.15cm}
\end{figure}

The input flows of an NTSS  constantly fluctuate. Hence, it is critical to be able to distinguish between robust and non-robust inputs. 
 The closer the robustness measures of inputs to the robustness threshold $\mathit{rbTh}$, the more non-robust those inputs are. Figure~\ref{fig:robustnessmeasurerange} illustrates the range of our robustness measure and specifies the robust and non-robust parts within this range. Test inputs whose robustness measures are close to $\mathit{rbTh}$ are more likely to flip system behaviour from being acceptable to unacceptable (or vice versa) by a small change, e.g., a change of \%1 in the flow bandwidths. Dually, inputs with robustness measures far away from $\mathit{rbTh}$ are unlikely to flip system behaviour due to minor perturbations in the input. 

\subsection{Non-Robust Behaviour Characterization}
\label{subsec:generation}
\begin{figure}[t]
    \centering
    \includegraphics[width=0.75\columnwidth]{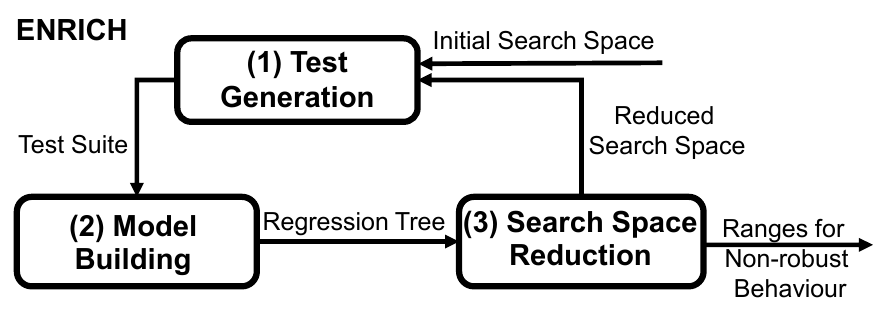}
    \imagecaptionspace
    \caption{An Overview of ENRICH.}\label{fig:approach}
    \vspace{-0.5cm}
\end{figure}

Figure~\ref{fig:approach} shows an overview of ENRICH --  our approach for characterizing non-robust test inputs. ENRICH performs, in an iterative manner, the following tasks: (1)~Generating a set of test inputs ($\mathit{TS}$) within a given search space.  (2)~Building a regression tree  model  ($\mathit{RT}$) using the test inputs $\mathit{TS}$ generated in the previous step and their robustness measure outputs. (3)~Using the regression tree model from step~2  to compute input ranges yielding non-robust outputs (i.e., robustness measures close to the robustness threshold $\mathit{rbTh}$). These ranges  are used as the reduced search space in the next iteration. The process continues until the test budget runs out.

\begin{figure}
    \centering
    \includegraphics[width=0.9\columnwidth]{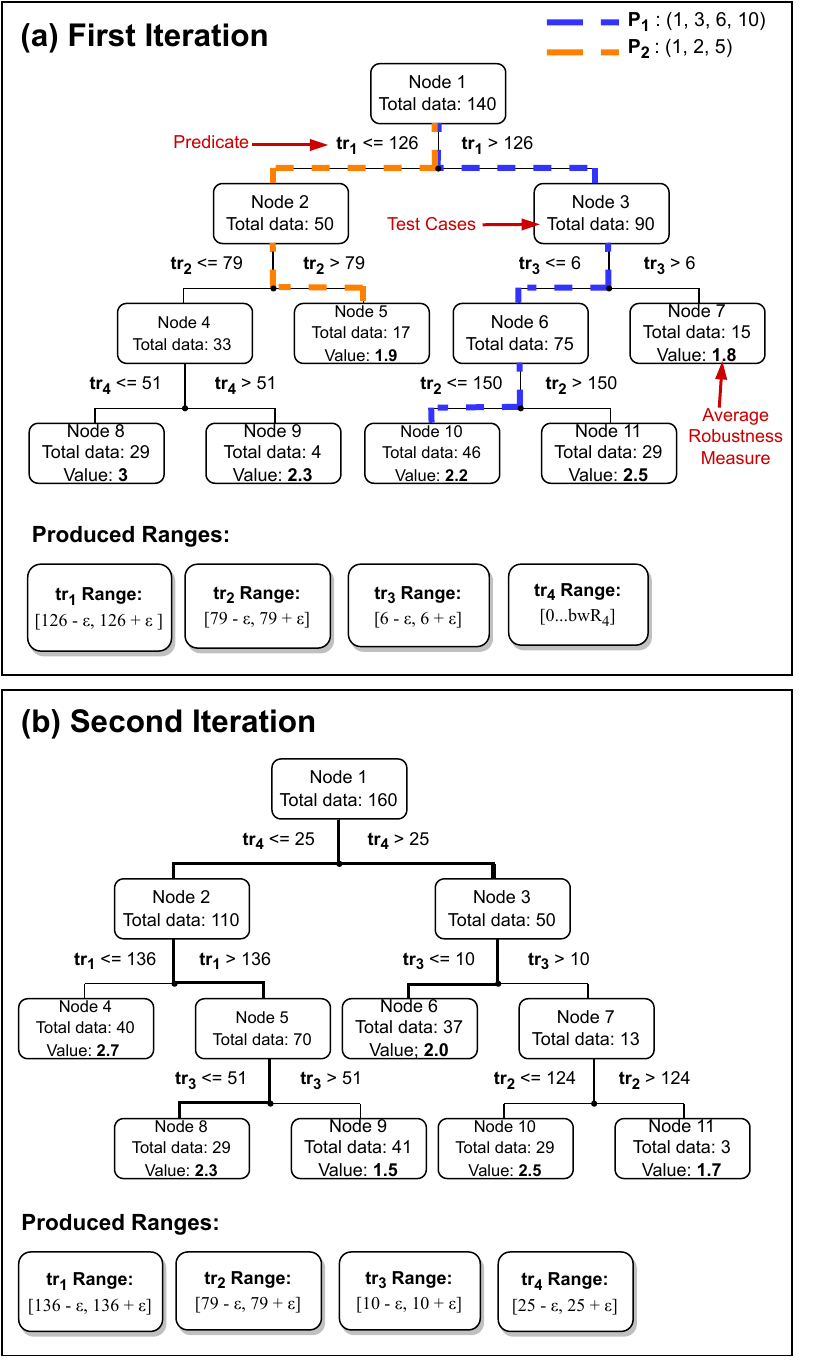}
    \caption{Illustration of two successive iterations of ENRICH and the induced input ranges.}
    \label{fig:decisiontree}
    \vspace*{-.5cm}
\end{figure}

Figure~\ref{fig:decisiontree} shows two regression trees generated by ENRICH in two consecutive iterations. The example is for an NTSS setup with four classes. Hence, each test input has four input variables $tr_1$ to $tr_4$. The initial search space for each $tr_i$ is  $\mathit{[0..\mathit{bwR}_i]}$.  ENRICH constrains the range for each $\mathit{tr}_i$ iteratively so that the range can capture  values leading to non-robustness. In the first iteration (Figure~\ref{fig:decisiontree}(a)), ENRICH is able to constrain ranges for $\mathit{tr}_1$, $\mathit{tr}_2$ and $\mathit{tr}_3$, but the range for $\mathit{tr}_4$ is left unconstrained. Note that the ranges generated by ENRICH are parameterized  in the form of $[v-\varepsilon, v+\varepsilon]$, where $v$ is a value derived from the regression tree representing the boundary between acceptable and unacceptable  values for the robustness measure. Later 
in this section, we describe in detail how the value $v$ for a variable $\mathit{tr}_i$ is derived from the regression tree. In the next iteration, for each variable $tr_i$ that is already constrained, we generate tests by sampling  $tr_i$ in $[v - $5$\%\cdot TB,  v + $5$\%\cdot TB]$. This is to ensure that ENRICH is  exploiting the search region that is in the close proximity of  the non-robustness threshold.

Figure \ref{fig:decisiontree}(b) shows the regression tree generated in the second iteration and the ranges induced by this tree. In the second iteration, we are able to constrain  $tr_4$, and refine the ranges for $tr_1$ and $tr_3$ into more precise ranges. We note that the new ranges for $tr_1$ and $tr_3$   are not too far away from their ranges in the previous iteration. That is, the search for the input ranges tends to exploit specific areas in the search space. We further note that in the second iteration, the tree does not produce any ranges for $tr_2$. In this case, ENRICH retains the range from the previous iteration, thus ensuring that 
input ranges are replaced only when a more precise range has been computed. In Section~\ref{sec:eval}, we evaluate the accuracy of
the final parameterized ranges generated by ENRICH to assess how well these ranges  capture non-robustness in NTSS. In particular, we study the relationship between the size of the ranges (i.e., the value of $\varepsilon$) and the accuracy of ENRICH.

ENRICH is implemented by Algorithm~\ref{alg:epicurus}. The input of ENRICH is a simulator,  $\mathit{Sys}$ (e.g., SOHOSim), and input ranges $R_1, \ldots, R_n$ (i.e., 
$\mathit{[0..\mathit{bwR}_1]} \times \ldots \times \mathit{[0..\mathit{bwR}_n]}$). ENRICH uses an adaptive random testing (ART)~\cite{metaheuristicsbook} algorithm to generate a set $\mathit{TS}$ of test cases (Line~3). ART randomly samples tests from the search space by maximizing the distance between newly selected vectors and the previously generated ones, hence ensuring that the tests are evenly spread over the search space.  ENRICH selects  each test case in $\mathit{TS}$ from the given input space $(R_1, \ldots, R_n)$ and executes it to compute its robustness measure.

\begin{minipage}{.9\columnwidth}
\vspace*{-.1cm}
\begin{algorithm}[H]
\scriptsize
\caption{The ENRICH Approach.}
\label{alg:epicurus}
\begin{flushleft}
\textbf{Inputs}. $\mathit{Sys}$: Executable system or simulator \hfill \\
 \hspace{0.75cm}  $(R_1, \ldots, R_n)$: The initial search space \\
 \textbf{Output}. NR: Input ranges characterizing non-robustness\\
 \end{flushleft}
\begin{algorithmic}[1]
\State \hspace{-0.1cm} TS $=$ TSAll $=\emptyset$; NR = $(R_1, \ldots, R_n)$;  \Comment{Variables Initialization}
\State \hspace{-0.1cm} \textbf{for} (i = 0 to $\textsc{Budget}$) \textbf{do:}
\State  \hspace{0.1cm} TS = \Call{GenTests}{$\mathit{Sys}$, NR} \Comment{Test input generation}
\State  \hspace{0.1cm} TSAll = TS $\cup$ TSAll; \Comment{Combine new and old tests}
\State  \hspace{0.1cm} $RT$ = \Call{BuildRT}{TSAll}; \hspace{-.3cm} \Comment{Build regression tree} \label{ln:dt}
\State \hspace{0.1cm} NR = \Call{Reduce}{$RT$}; \Comment{Search space reduction} \label{ln:bc}
\State \hspace{-0.1cm} \textbf{end for};
\State \hspace{-0.1cm} \Return NR;
\end{algorithmic}
\end{algorithm}
\vspace*{.05cm}
\end{minipage}

ENRICH builds a regression tree model using the test suite built so far (Lines~4-5). See Figure~\ref{fig:decisiontree} for examples of regression trees built by ENRICH. A regression tree recursively splits data at each inner node into two children by minimizing the sum of the squared deviations from the mean in each child node~\cite{regressiontree}.
The value of each leaf node represents the average of the robustness measures of the test cases in that node. The tree edges are labelled with predicates in the form of $tr_i \sim v$ where $tr_i$ is an input variable, $v \in \mathbb{R}$ and $ \sim\ \in \{\leq, >\}$.

Finally, ENRICH uses the regression tree model to constrain the ranges of the input variables (Line 6). A description of the $\textsc{Reduce()}$ routine is available in our supplementary material~\cite{algsandfigs}. Briefly, ENRICH finds the two paths whose leaf-node values, i.e., the average of the robustness measure, are closest to the robustness measure threshold ($\mathit{rbTh}$). In  Figure~\ref{fig:decisiontree}(a), assuming that $\mathit{rbTh} = 2.1$, we select \hbox{$P_1\!:$\,(1, 3, 6, 10)} and $P_2\!:\,$(1, 2, 5). We identify constraints 
induced by these paths such that the constraints  only characterize test cases with robustness measures close to the threshold. In Figure~\ref{fig:decisiontree}(a), the constraint induced by  $P_1$ (the path highlighted in blue) is ($tr_1 > 126) \land (tr_3 \leq 6) \land (tr_2 \leq 150 $), and the constraint induced by  $P_2$ (the path highlighted in orange) is ($tr_1 \leq 126) \land (tr_2 > 79$).
The predicate $tr_2 \leq 150$ is not contributing towards identifying  non-robust behaviours because its related branch at node 6 splits node 10 and 11 whose values are both above $\mathit{rbTh}$.  Therefore, we eliminate the predicate $tr_2 \leq 150$.  Specifically, we keep the predicates that lead to two nodes where one is above and one is below $\mathit{rbTh}$, and remove the rest as they do not help with the characterization of non-robust behaviours. We then simplify the constraints so that for each variable $tr_i$, we obtain at most one upper-bound predicate ($tr_i \leq v$) and one lower-bound predicate ($tr_i > v$), since, for example,  any two predicates  ($tr_i \leq v_1$) and  ($tr_i \leq v_2$) such that  ($v_1 \leq v_2$) can be replaced by ($tr_i \leq v_1$). At the end, for each predicate $tr_i \sim v$ where $\sim\ \in \{\leq, >\}$, we create a (parameterized) range $[\mathit{max}(v - \varepsilon, 0), \mathit{min}(v + \varepsilon, \mathit{bwR}_i)]$ where ${bwR}_i$ is the maximum value that $tr_i$ can assume. %We discuss in Section~\ref{sec:eval} how the $\varepsilon$  parameter  can be set to support two different use cases: the more accurate generation (prediction) of non-robustness versus more accurate characterization (coverage) of non-robustness.
For each $\mathit{tr_i}$, its new range is passed to the test generation routine (Line 3).
In the test generation routine, each $tr_i$ for which  a range $[\mathit{max}(v - \varepsilon, 0), \mathit{min}(v + \varepsilon, \mathit{bwR}_i)]$ exists is sampled  close to $v$ (within $v\pm 5\% \cdot TB$).  If $\mathit{tr_i}$ is not constrained, then its range obtained in the previous iteration will be retained for the next iteration. For each variable $tr_i$, ENRICH ensures that  the default range of  $\mathit{[0.. \mathit{bwR}_i]}$ will not be passed to the next iteration if the variable's range has been narrowed at some point. This is essential as the ultimate goal of the algorithm is to get narrower ranges for each $tr_i$.

%Finally, we extract ranges for the input variables using the constraints induced by  the paths $P_i$ and $P_k$ (line~4 of Algorithm~\ref{alg:reduction}).  Note that, for each variable $tr_i$, we obtain at most one upper-bound predicate ($tr_i \leq v$) and one lower-bound predicate ($tr_i > v$), since, for example,  any two predicates  ($tr_i \leq v_1$) and  ($tr_i \leq v_2$) such that  ($v_1 \leq v_2$) can be replaced by ($tr_i \leq v_1$). We simplify the lower-bound predicates in a similar fashion. For each predicate $tr_i \sim v$ where $\sim \in \{\leq, >\}$, we create a range $[\mathit{max}(v - \varepsilon, 0), \mathit{min}(v + \varepsilon, \mathit{bwR}_i)]$ where ${bwR}_i$ is the maximum value that $tr_i$ can assume, and $\varepsilon$ is a parameter of our algorithm (see Table~\ref{tab:options}). %For any  $tr_i$ that is not constrained by any predicate, we keep the default range ($\mathit{[0.. \mathit{bwR}_i]}$). 

%For example, in Figure~\ref{fig:decisiontree}(a),  we extract the following predicates from $P_1$ and $P_2$: $126 < \mathit{tr_1} \leq 126$, $79 < \mathit{tr_2}$ and $\mathit{tr_3} \leq 6$. Assuming that $\mathit{bwR_1} = 400$, $\mathit{bwR_2} = 370$, $\mathit{bwR_3} = 100$, and $\varepsilon = 20$ (i.e., $5\%$ of the highest bandwidth), we obtain the ranges $[106, 146]$, $[59, 99]$ and $[0, 26]$ for $tr_1$, $\mathit{tr_2}$, and $\mathit{tr_3}$, respectively. 

\section{Evaluation}
\label{sec:eval}
We evaluate ENRICH and SOHOSim using two RQs:

\textbf{RQ1. (Accuracy of ENRICH)} \emph{Do the ranges generated by ENRICH accurately capture  non-robust behaviours of an NTSS?} We develop a set of tests labelled as robust and non-robust. We then use the ranges produced by ENRICH to predict the labels for these tests and assess the prediction accuracy of ENRICH. Since the ranges generated by ENRICH are parameterized (i.e., by the $\varepsilon$ parameter), we evaluate the accuracy of ENRICH by varying $\varepsilon$ and discuss how $\varepsilon$ can be set in practice to address different needs: more accurate generation of non-robustness versus more accurate characterization (or coverage) of non-robustness. Further, since ENRICH is randomized, we study whether combining ranges obtained from multiple runs of ENRICH improves its accuracy.

\textbf{RQ2. (Accuracy of SOHOSim)} \emph{Is there a significant difference between the test results obtained from SOHOSim (our simulator) and a hardware-in-the-loop testbed of NTSS?} Network testing is often performed using Hardware-in-the-Loop (HiL) testbeds~\cite{himmler2013hardware}. SOHOSim, being a virtual testbed, provides flexibility and efficiency, allowing us to run a large number of tests from which we can infer interpretable feedback (i.e., input ranges characterizing non-robustness). Nevertheless, as this RQ investigates, we need to ensure that the simulation results are close to those obtained over an actual NTSS that executes on hardware. To assess the accuracy of SOHOSim, we compare the  results obtained from SOHOSim with the results obtained from a HiL-based NTSS testbed with identical configurations.

\textbf{Data Availability.} We have made the installation guidelines for SOHOSim, our implementations of ENRICH and BASELINE, and our experimental results publicly available~\cite{enrichgithub}.

\subsection{RQ1-Accuracy of ENRICH}
\label{sec:experimentdesign}
Before answering RQ1, we present the baseline, our experiments' parameters and setup, and the comparison metrics.

%To our knowledge, our work is the first to apply automated software testing to NTSS and to use the test results to infer ranges characterizing non-robustness in NTSS.

\textbf{Baseline:} To the best of our knowledge, there is no approach in the literature that performs what ENRICH does for network traffic-shaping systems. While there are approaches that develop interpretable ML from test results~\cite{raja2018, gopinath2020abstracting}, none generate constraints in the form of ranges for input variables. To have a baseline, as per the empirical guidelines for search-based software engineering~\cite{guidelines},  we compare our approach with standard adaptive random testing (ART) and use the results to infer input ranges. In particular, our baseline (hereafter BASELINE) generates tests using ART within the default input ranges and then builds a regression tree using the test results only once. BASELINE uses the same parameters as ENRICH's, the main difference being that BASELINE is non-iterative. That is, it obtains (parameterized) input ranges based on the results of a fully explorative search and skips the  combination of explorative and exploitative searches as utilized by ENRICH. 

\textbf{Parameters and setup:} The experiment parameters  are shown in Table~\ref{tab:options}.  We configure SOHOSim (a conceptualization of which was provided in Figure~\ref{fig:virtualsetup}) according to an industrial setup of NTSS recommended by RabbitRun. We refer to this setup as \emph{NTSS-RR}. This setup uses an \textit{8-tier mode} of CAKE known as diffserv8~\cite{cakepaper}, (i.e., $n=8$). We use the default configuration of CAKE for the maximum bandwidth ($\mathit{bwR_i}$) of each class. We set the total bandwidth  ($\mathit{TB}$ in Table~\ref{tab:options}) to 400 Mbit as per the recommendation of our partner.

\begin{table}[t]
\caption{The parameters required by ENRICH}
\label{tab:options}
\vspace{-.15cm}
\scalebox{0.7}{
\begin{tabular}{|l|p{8cm}|c|}
\hline
\multicolumn{1}{|c|}{\textbf{Parameter}} & \multicolumn{1}{c|}{\textbf{Definition}}                         & \textbf{Value} \\ \hline
$\mathit{rbTh}$                                 &  The robustness measure threshold                                  & 3.6            \\ \hline
$\mathit{TestSuiteSize}$                        & Number of test cases generated in each iteration of ENRICH       & 20             \\ \hline
$\mathit{NodeSize}$                             & Minimum number of tests at the leaves of a regression tree      & 1              \\ \hline
%\textcolor{blue}{$\mathit{Epsilon} (\varepsilon)$}                              & The offset specifying the range width from which we sample data in each iteration of ENRICH                     & 5\%             \\ \hline
$\mathit{TB}$                       &   
The NTSS total bandwidth   & 400         \\ \hline
\end{tabular}}
\vspace{-.3cm}
\end{table}

%\emph{Note that the final ranges obtained from ENRICH  remain to be parameterized by $\varepsilon$. This value is only used by the internal iterations of ENRICH to guide an exploitative search.}  

To answer RQ1, we apply ENRICH and BASELINE to NTSS-RR.  We set the total number of tests generated and simulated by each run of ENRICH and BASELINE to $300$. We arrived at this number based on preliminary experiments and setting a time budget of one day (give or take) for one run. In our  setup, it takes approximately 27 hours, on average, for SOHOSim to perform $300$ simulations. 

For each run of ENRICH, in the first iteration, we generate $100$ test inputs to have a sizeable number of data points for building an initial regression tree. We then perform $10$ iterations of Algorithm~\ref{alg:epicurus} where we generate $20$ test inputs in each iteration, thus ensuring that the total number of simulations by ENRICH is $300$. 
The stopping criterion ($\mathit{NodeSize}$) for regression tree creation is one, meaning that we expand the tree to the fullest extent. This makes it possible to derive constraints from the tree involving the most input variables, and hence, obtain more constrained input ranges. Based on feedback from our  partner, we set the robustness threshold $\mathit{rbTh}$ to $3.6$.  We run ENRICH and BASELINE fifteen times each to account for random variation. Collectively, it took more than three weeks of computation to carry out all the runs. Performing more runs was not feasible due to time limitations.  All the experiments were executed on a machine with a 2.5 GHz Intel Core i9-119900H CPU and 64~GB of DDR4 memory. 

\textbf{Comparison Metrics.} We generate a set $\mathit{TestSet}$ of $200$ test inputs randomly and label them as robust and non-robust using the following procedure: For each test $\mathit{tc} \in \mathit{TestSet}$, if the calculated robustness measure is below (resp. above) $\mathit{rbTh}$, we deduct (resp. add) a small perturbation over a short time period (around $2$\% of $\mathit{TB}$) from (resp. to) each input value in $\mathit{tc}$ and simulate the resulting test to check whether the robustness measure has moved from below $\mathit{rbTh}$ to above it, or vice versa. A test input is labelled robust, if its robustness measure does not move from below $\mathit{rbTh}$ to above it, or vice versa. Otherwise, the test input is labelled as non-robust. Labelling test inputs as robust and non-robust is expensive since we need to run each test multiple times. It took more than two days to generate a labelled set of $230$ test inputs. The perturbation size ($2$\%) is based on the recommendations of our partner.

We label the tests in $\mathit{TestSet}$ based on the input ranges that ENRICH and BASELINE generate. For example, if one run of ENRICH (or BASELINE) generates ranges $[180..280]$ for $tr_1$ and $[32..42]$ for $tr_2$, we label a test as non-robust if and only if the values of both $tr_1$ and $tr_2$ in that test fall in those ranges. Otherwise, we label the test as robust. 

\begin{figure}
    \centering
    \includegraphics[width=\columnwidth]{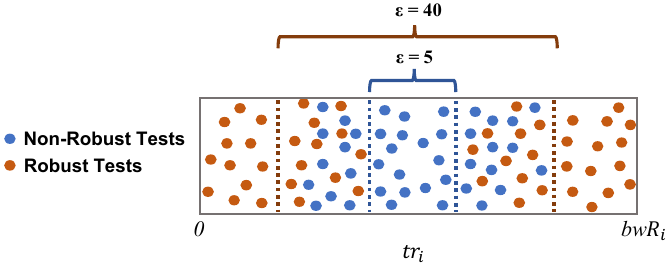}
    \caption{Distribution of robust and non-robust NTSS tests with respect to the default range of an input variable $\mathit{tr}_i$. Smaller ranges (lower $\varepsilon$ values) are more precise in predicting non-robustness, and larger ranges (higher $\varepsilon$ values) provide more coverage for characterizing non-robustness.}
    %the input ranges induced by the second regression tree.}
    \label{fig:epsilon}
    \vspace*{-.5cm}
\end{figure}

We use the accuracy metric to compare the prediction ability of ENRICH and BASELINE. Accuracy is the number of correctly predicted tests over the total number of tests.  The accuracy metric is a preferred single-number evaluation metric~\cite{machinelearningyearning} that can assess the correctness of a technique for predicted classes (i.e., robustness and non-robustness in our case). 
As noted earlier, the ranges generated by ENRICH and BASELINE for each input variable are in the form of $[v-\varepsilon, v+\varepsilon]$. We thus need to assign a value to $\varepsilon$ to use the ranges for labelling. The value of $\varepsilon$ allows us to control whether the input ranges are good at accurately generating non-robustness or at covering (characterizing) non-robustness.
%Setting $\varepsilon$ to a small value (e.g., $5$\% of the default range of an input variable) leads to a narrow range, increasing the accuracy for the prediction of non-robustness \textcolor{blue}{resulting in higher accuracy}. In contrast, setting $\varepsilon$ to a large value (e.g., $20$\%  to $40$\% of the default range) leads to a wide range, increasing the precision for the prediction of robustness \textcolor{blue}{which results in higher overall accuracy}. For example, Figure~\ref {fig:epsilon} shows setting $\varepsilon$ to $5$\% versus $40$\% of the default range of a variable makes an input range more precise for predicting non-robustness versus robustness, respectively. Intuitively, a narrow range ($\varepsilon = 5$\%) is precise for the non-robustness class and can also prune robust tests well (i.e., it has a high recall for the robustness class). Dually, a wide range ($\varepsilon = 40$\%) is more precise for the robustness class with a high recall for the non-robust cases.
%%%BAHARIN:
Figure~\ref {fig:epsilon} is a schematic view of the
distribution of robust and non-robust tests for NTSS when the value of an input variable $\mathit{tr}_i$ changes in its default bandwidth range $[0..\mathit{bwR}_i]$. When the values of input variables are close to the lower or upper bounds of their ranges, the resulting tests are more likely to be robust (i.e., robustly good for the lower bound, and robustly bad for the upper bound). But, when the values of input variables are in the middle, the tests are more likely to be non-robust. The input ranges generated in our work, which often fall in the middle, can be used for two different use cases: (1) They can be used to precisely predict (generate) non-robustness (i.e., yielding high precision for non-robustness), or (2) They can be used to characterize (cover) non-robustness (i.e., yielding high recall for non-robustness). As shown in Figure~\ref {fig:epsilon}, we expect the ranges obtained by setting $\varepsilon$ to a small value (e.g., $5$\% of the default range) to be better at predicting non-robustness, and the wider ranges (e.g., $\varepsilon = 40$\%) to be better at covering (characterizing) non-robustness. Hence, in addition to accuracy, we report for smaller ranges the \emph{precision} for non-robustness, and for the wider ranges the \emph{recall} for non-robustness. Full precision and recall results for both robustness and non-robustness are available online~\cite{enrichgithub}.

Since ENRICH and BASELINE are randomized, they likely generate different results when they are executed multiple times. We expect ENRICH, but not BASELINE, to generate overlapping ranges over multiple runs and exploit specific sub-ranges of the default input ranges instead of exploring the entire default ranges. To assess this difference between ENRICH and BASELINE, we compare ENRICH and BASELINE based on both their single runs and combinations of their multiple runs. Specifically, provided with $n$ runs of ENRICH (resp. BASELINE),  we label a test in $\mathit{TestSet}$ as non-robust if and only if at least one of the $n$ runs label the test as non-robust. Otherwise, we label the test as robust. We vary $n$ from $1$ to $15$ (i.e., the maximum number of runs we have for both techniques). To account for randomness in selecting multiple runs, we randomly select $20$ different combinations for $n=2,\ldots, 13$. For $n=1, 14$, we consider all the $15$ possible combinations (${15}\choose{1}$ and ${15}\choose{14}$); and, for $n=15$, we consider all the runs together (${15}\choose{15}$). %We then report the accuracy and recall values for all the $n$ combinations selected for ENRICH and BASELINE to study the accuracy and convergence of these techniques by considering the combinations of the results from their multiple runs. 

%To answer RQ1, we compare ENRICH and BASELINE based on the accuracy metric for different values of $\varepsilon$, since we expect ENRICH to always be more accurate than BASELINE. For narrow ranges ($\varepsilon$ = $5$\%), we are interested in recall for robustness, since we expect such ranges to be used for characterizing robustness (i.e., to be more precise at generating non-robustness). Dually, for wide ranges ($\varepsilon$ = $25$\% to $40$\%), we are interested in recall for non-robustness, since we use these ranges to characterize non-robustness (i.e., to be more precise at generating robustness).

\textbf{Results.} To answer RQ1, we show the results for two cases: (1)~Input ranges are used to generate non-robustness when we set $\varepsilon$ = $5$\% to have narrow ranges, and (2)~Input ranges are used to characterize (cover) non-robustness when we set $\varepsilon$ = $25$\% to $40$\% to have wide ranges. As noted earlier, for both cases, we report  accuracy. In addition, we  report the precision for the non-robust class  to assess the generation of non-robust behaviours (case-1), and the recall for the non-robust class to assess the coverage of non-robust behaviours (case-2). 

\begin{figure}
    \centering
    \begin{subfigure}[]{\columnwidth}
        \includegraphics[width=\columnwidth]{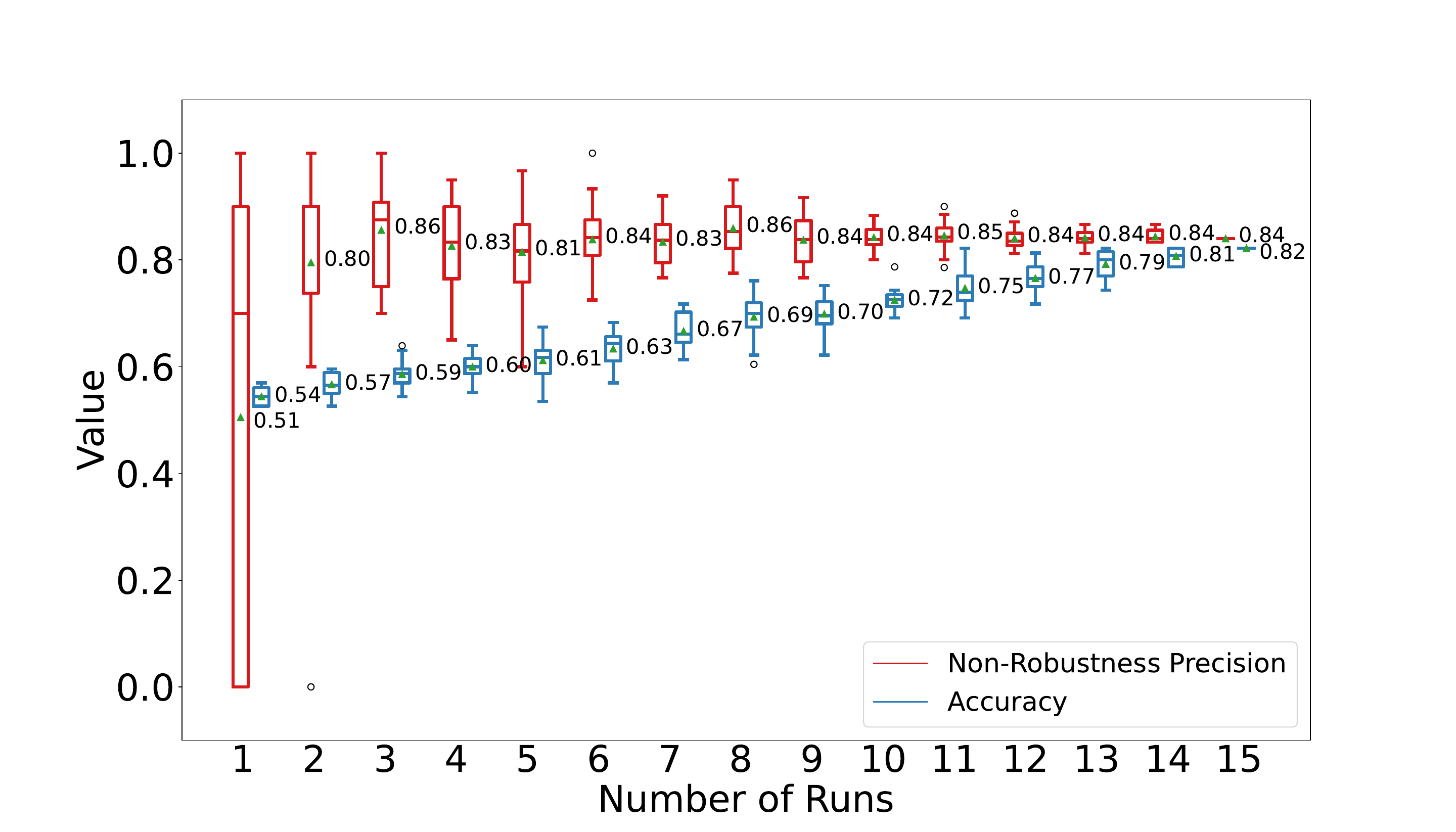}
        \caption{Accuracy and non-robustness precision for ENRICH.}
        \label{fig:5NRP}
    \end{subfigure}%
    
    ~ %add desired spacing between images, e. g. ~, \quad, \qquad etc.
    \begin{subfigure}[]{\columnwidth}
        \includegraphics[width=\columnwidth]{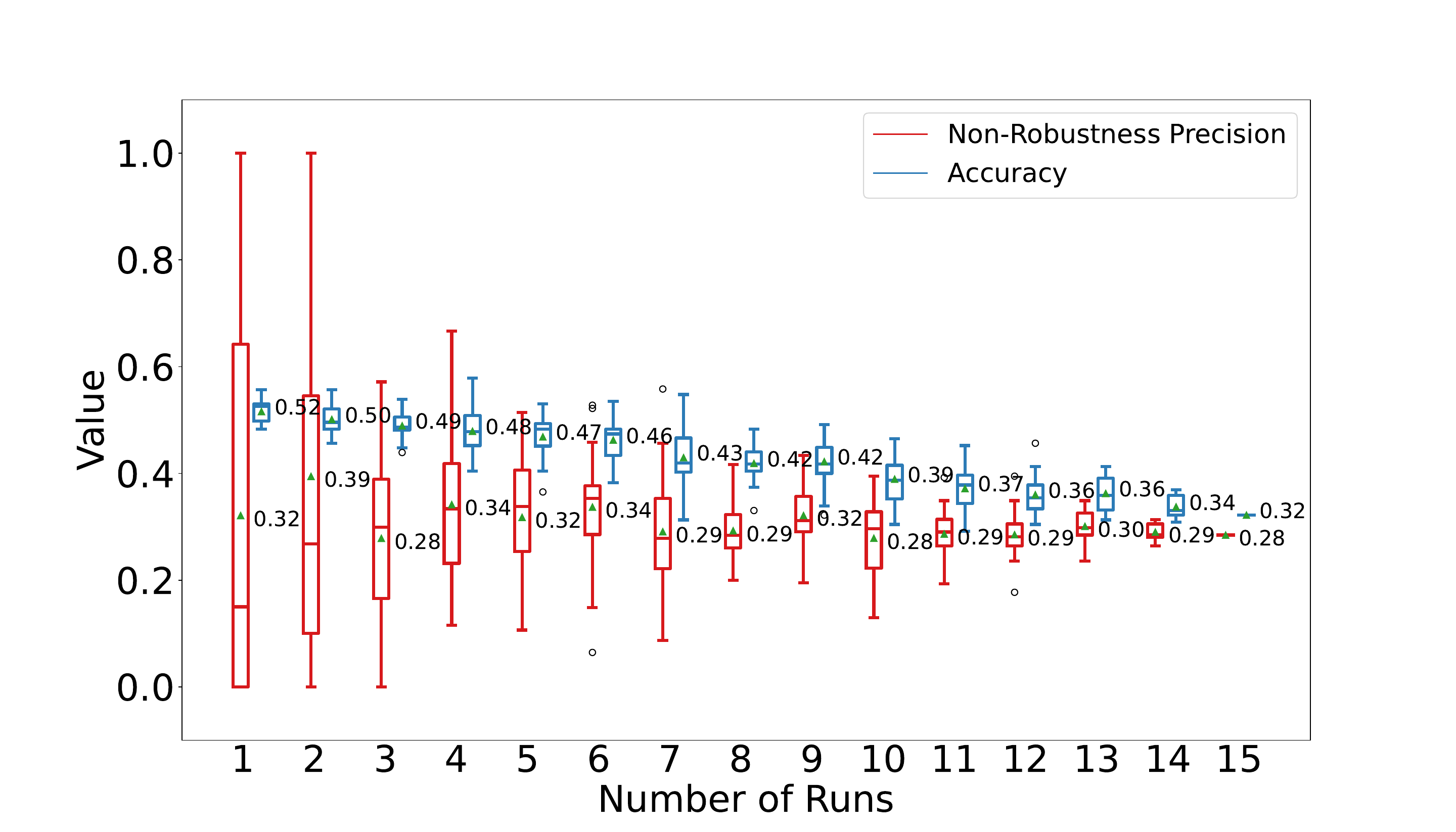}
        \caption{Accuracy and non-robustness precision for BASELINE.}
        \label{fig:5NRPbaseline}
    \end{subfigure}
    ~ %add desired spacing between images, e. g. ~, \quad, \qquad etc.
    
    \caption{Evaluating the non-robustness generation use case (\hbox{$\varepsilon = 5\%$}) by comparing accuracy and non-robustness precision for different run combinations of ENRICH and BASELINE.}\label{fig:5NRRP}
    
\end{figure}

Figure~\ref{fig:5NRRP} shows the results for the first case which include the accuracy and  non-robustness precision obtained from the combinations of $n$ random runs of ENRICH and BASELINE, where we vary $n$ from $1$ to $15$.  As shown in the figure, the average accuracy and non-robustness precision of ENRICH are always higher than those of BASELINE. As we consider more run combinations, both the accuracy and the non-robustness precision of ENRICH increase significantly (from an average of $54$\% to $82$\% for  accuracy, and from $51$\% to $84$\% for non-robustness precision). For BASELINE, however, both accuracy and non-robustness precision decrease considerably (from an average of 52\% to 32\% and from an average of 32\% to 28\%, respectively) as we consider more run combinations. Overall, using ENRICH, we are able to obtain an average accuracy of $82$\% and an average non-robustness precision of $84$\% when we consider all the runs. In contrast, the best averages for accuracy and non-robustness precision we  obtain using BASELINE are $52$\% and $39$\%, respectively; these results are significantly lower than ENRICH's.

%for all the combinations, ENRICH yields an average recall for robustness higher than $87$\% and an average \textcolor{blue}{accuracy} higher than \textcolor{blue}{$60$\% (except for the case of single, two and three runs)}. The results suggest that by combining the results from multiple runs of ENRICH, we improve the overall accuracy of non-robustness prediction (or generation).  For example, using $10$ runs of ENRICH, the \textcolor{blue}{accuracy} is \emph{always} above \textcolor{blue}{$70$\% and the recall is \emph{always} above $87$\%}. On the other hand, as shown in Figure~\ref{fig:5NRRP}(b), the average \textcolor{blue}{accuracy} of BASELINE is consistently lower than \textcolor{blue}{$52$\%} for all the run combinations, and its average recall (significantly) drops from $96$\% to $36$\% as we combine the results from more runs. Overall, the \textcolor{blue}{accuracy} and recall values for \textcolor{blue}{non robustness prediction} of ENRICH are significantly higher than those obtained from BASELINE. \textbf{statistical tests?}

\begin{figure}
\vspace{-0.52cm}
    \centering
    \begin{subfigure}[]{\columnwidth}
        \includegraphics[width=\columnwidth]{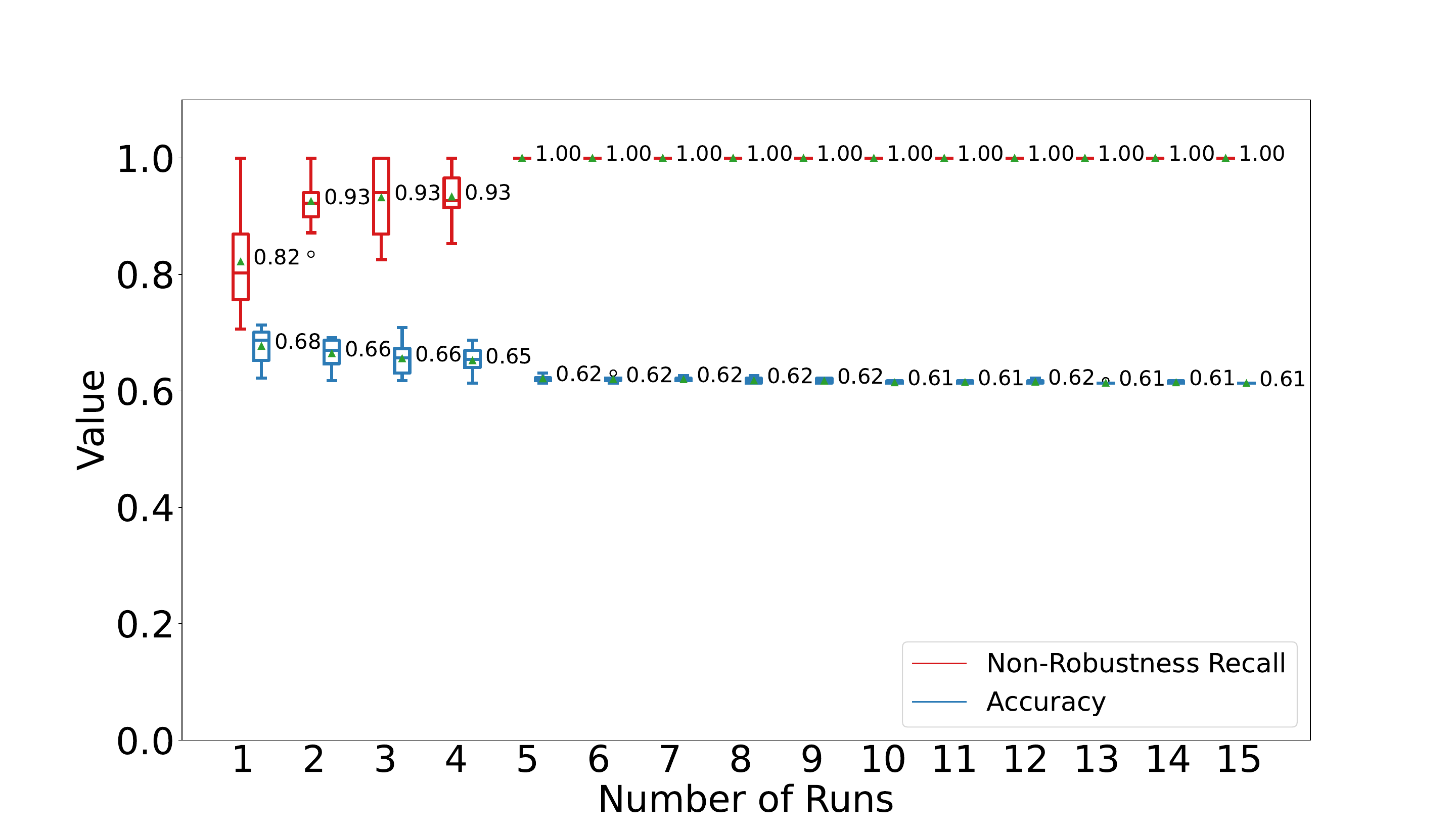}
        \caption{Accuracy and Non-Robustness Recall of ENRICH.}
        %\label{fig:5NRP}
    \end{subfigure}%
    
    ~ %add desired spacing between images, e. g. ~, \quad, \qquad etc.
    \begin{subfigure}[]{\columnwidth}
        \includegraphics[width=\columnwidth]{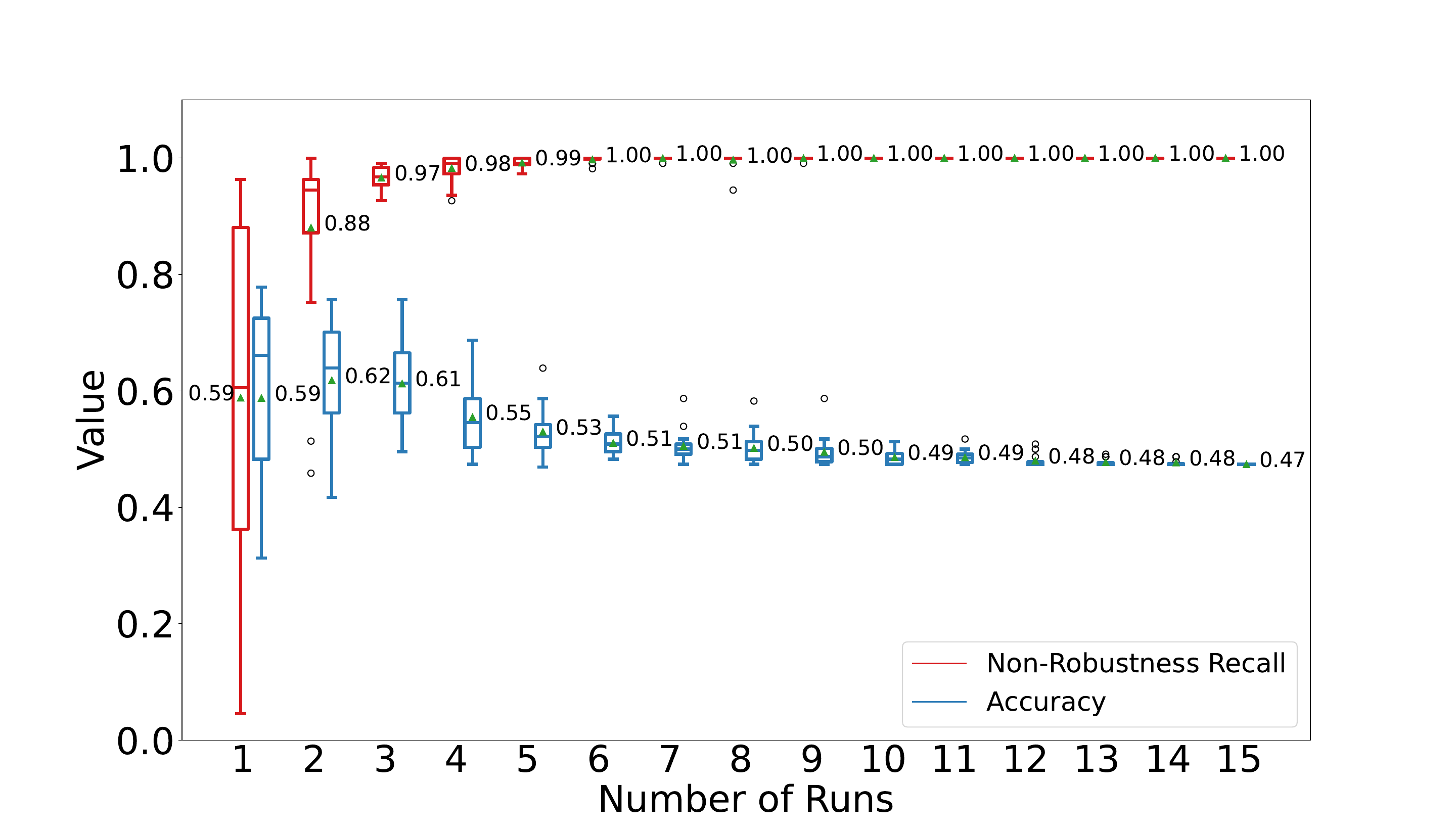}
        \caption{Accuracy and Non-Robustness Recall of BASELINE.}
        %\label{fig:5NRPbaseline}
    \end{subfigure}
    ~ %add desired spacing between images, e. g. ~, \quad, \qquad etc.
    
    \caption{Non-robustness characterization use case (\hbox{$\varepsilon = 40\%$});  comparing accuracy and non-robustness recall for different run combinations of ENRICH and BASELINE.}\label{fig:25to40}
    \vspace{-0.52cm}
\end{figure}

Dually to Figure~\ref{fig:5NRRP}, Figure~\ref{fig:25to40} compares ENRICH and BASELINE for the second case when we have  $\varepsilon = 40$\% of the default ranges.  The figure compares the accuracy and  non-robustness recall for the $n$ run combinations of ENRICH and BASELINE where $n$ varies from $1$ to $15$. Similar to the previous case, the average accuracy of ENRICH is always higher than that of BASELINE. The average non-robustness recall of ENRICH is higher than that of BASELINE for a single run and is almost the same for the other run combinations. Both approaches achieve almost $100$\% non-robustness recall for $n \geq 6$. However, when both approaches achieve full recall, ENRICH always maintains an average accuracy that is  $10$\% or more higher than that of BASELINE. In addition, we show the averages for accuracy and non-robustness recall
for  $\varepsilon$ = $25$\%, $30$\% and $35$\%  and for $n=1, 5, 10, 15$ in Table~\ref{tbl:avgbox}. The full box-plots are available online~\cite{algsandfigs}. The results in the table are consistent with those for $\varepsilon$ = $40$\%. Specifically, ENRICH is always more accurate than BASELINE. For the $10$- and $15$-run combinations where both approaches achieve a non-robustness recall of above $90$\%, ENRICH maintains an average accuracy that is $13$\% to $24$\%  higher than that of BASELINE.

\begin{table*}[t]
	\caption{Comparison of averages of  accuracy and  recall for the non-robustness class of ENRICH (E) and BASELINE (B) at $\varepsilon = 25\%, 30\%$ and $35\%$, and for their single  runs as well as  combinations of their $5$, $10$ and $15$ runs.}
	\label{tbl:avgbox}
	\vspace*{-.1cm}
	\centering
	\scalebox{0.65}{
	\begin{tabular}{|c|c|c||c|c||c|c||c|c|}
	\toprule
	  &  
	  \multicolumn{2}{c||}{ \textbf{\emph{$1$ Run}}} & \multicolumn{2}{c||}{\textbf{\emph{$5$ Runs}}} & \multicolumn{2}{c||}{\textbf{\emph{$10$ Runs}}} & \multicolumn{2}{c|}{\textbf{\emph{$15$ Runs}}} \\
	 \hline
 & \textbf{Accuracy(E - B)}& \textbf{Non-Robustness Recall(E - B)} & \textbf{Accuracy(E - B)}& \textbf{Non-Robustness Recall(E - B)} & \textbf{Accuracy(E - B)}& \textbf{Non-Robustness Recall(E - B)} & \textbf{Accuracy(E - B)}& \textbf{Non-Robustness Recall(E - B)}   \\
 \hline
$\varepsilon = 25\%$ & 
$0.64$ -- $0.61$ & $0.42$ -- $0.53$ & $0.72$ -- $0.59$ & $0.87$ -- $0.98$ & $0.72$ -- $0.52$ & $0.93$ -- $1.00$ & $0.71$ -- $0.47$ & $0.95$ -- $1.00$ 
 \\

$\varepsilon = 30\%$ &  
$0.68$ -- $0.61$ & $0.65$ -- $0.55$ & $0.70$ -- $0.55$ & $0.90$ -- $0.98$ & $0.69$ -- $0.49$ & $0.94$ -- $1.00$ & $0.68$ -- $0.47$ & $0.97$ -- $1.00$ 
 \\

$\varepsilon = 35\%$ & 
$0.66$ -- $0.64$ & $0.71$ -- $0.60$ & $0.64$ -- $0.56$ & $0.94$ -- $0.99$ & $0.63$ -- $0.50$ & $0.96$ -- $1.00$ & $0.62$ -- $0.47$ & $0.98$ -- $1.00$
 \\ 
\bottomrule
\end{tabular}}

\vspace*{.1cm}	
%{\scriptsize \hspace{.5cm} * The Precision(E - B) and Recall(E -B) columns show the Precision and Recall metrics for Enrich (E) versus those for BASELINE (B).\hfill\mbox{}}
\vspace*{-.35cm}
\end{table*}

Overall, the results show that the ranges generated by ENRICH, due to its exploitative search, tend to converge in a way that they can generate and characterize non-robustness with a higher accuracy compared to BASELINE. On the other hand, the relatively low accuracy of BASELINE indicates that the non-robustness recall results of BASELINE are more due to the random (explorative) nature of its search. Specifically, the high recall for non-robustness in Figure~\ref{fig:25to40}(b) is because BASELINE labels \emph{all} the tests as non-robust. Hence, while it achieves high non-robustness recall, its accuracy is lower than $51$\%. This is confirmed by the detailed precision and recall results for both robustness and non-robustness that is available online~\cite{enrichgithub}.

\begin{mdframed}
\textit{The answer to \textbf{RQ1}} is that ENRICH consistently generates and characterizes non-robustness with a significantly higher accuracy than BASELINE. Specifically, ENRICH generates non-robustness with a precision of $84$\% while outperforming BASELINE in the overall accuracy by at least $30$\%. In addition, ENRICH characterizes non-robustness with $100$\% recall and yields an accuracy that is at least $10$\% higher than that of BASELINE. 
\end{mdframed}

\subsection{RQ2-Accuracy of SOHOSim} 
To answer \textbf{RQ2}, we randomly generate 100 tests, and simulate them on SOHOSim and on a hardware-in-the-loop (HiL) testbed for NTSS where we replace the Internet modem in Figure~\ref{fig:architecturegoals} with hardware. We refer to this testbed as  SOHOHW. 

We compute the robustness measures for each test based on the results obtained from SOHOSim and SOHOHW. Table~\ref{tbl:environment} shows the values of mean and standard deviation obtained for 100 test inputs executed on SOHOSim and SOHOHW.  We have conducted Mann–Whitney U Test (level of significance is $0.05$) on these two distributions. The p-value is 0.18 indicating that there is no significant difference between the two distributions. In addition, the mean absolute error (MAE) of the robustness measures of the 100 test inputs obtained from SOHOSim and SOHOHW is $0.25$, which is about $3\%$ of the robustness-measure range, noting that the range of the robustness measure is $[0.5..8.0]$. These results show that SOHOSim is an accurate simulator and a good proxy for a physical testbed. 

\begin{mdframed}[nobreak=true]
\textit{The answer to} \textbf{RQ2} is that there is no statistically significant difference between SOHOSim and SOHOHW. The mean absolute error between the robustness measure values obtained from SOHOSim and SOHOHW is $3$\%.
\end{mdframed}

\begin{table}[t]
\caption{Mean and standard deviation for random tests obtained from SOHOSim and SOHOHW. Two environments are also compared based on p-value and MAE.}
\label{tbl:environment}
\scalebox{0.85}{
\begin{tabular}{|cc|cc|cc|}
\hline
\multicolumn{2}{|c|}{\textit{\textbf{SOHOSim}}}                   & \multicolumn{2}{c|}{\textit{\textbf{SOHOHW}}}                    & \multicolumn{2}{c|}{\textit{\textbf{Comparison}}}    \\ \hline
\multicolumn{1}{|c|}{\textbf{Mean}} & \textbf{Standard Deviation} & \multicolumn{1}{c|}{\textbf{Mean}} & \textbf{Standard Deviation} & \multicolumn{1}{c|}{\textbf{P-value}} & \textbf{MAE} \\ \hline
\multicolumn{1}{|c|}{2.26}          & 0.995                       & \multicolumn{1}{c|}{2.48}          & 0.944                       & \multicolumn{1}{c|}{0.18}             & 0.25         \\ \hline
\end{tabular}}
\end{table}

%\begin{figure}[t]
%    \centering
%    \includegraphics[width=\columnwidth]{figs/omni-switchwithmeanvalues.pdf}
%    \caption{\textcolor{blue}{Results of RQ2 - Comparison of robustness measures obtained by SOHOSim and SOHO-HW}}\label{fig:Q3}
%    \vspace*{-.4cm}
%\end{figure}

\textbf{Threats to Validity.} Construct and external validity are the validity aspects most relevant to our evaluation.

\emph{Construct validity:} 
The main consideration in relation to construct validity is the degree of accuracy of our simulator, SOHOSim. The risk of SOHOSim not being representative of the real world is mitigated by RQ2, where we show that SOHOSim behaves similarly to a physical testbed.

\emph{External validity:} While our evaluation of ENRICH is based on a single case study, our experimental setup captures the most common situation in SOHO, where the office is connected to the Internet via a single traffic-shaping-enabled router. To further improve the external validity of our evaluation, we need to perform additional case studies involving network testbeds different than the one in our current case study and examine how effective ENRICH is at identifying non-robust input regions in different testbeds.

%We have strived to provide a precise formalization of NTSS upon which our approach is built upon. ENRICH can be applied to any NTSS with different settings (i.e., different number of classes and total bandwidth).
\vspace*{-.1cm}
\section{Related Work}
\label{sec:related}
\vspace*{-.05cm}
%We compare our approach with related work in robustness analysis and network testing. 

%	Justyna Petke, David Clark, William B. Langdon: Software robustness: a survey, a theory, and prospects. ESEC/SIGSOFT FSE 2021: 1475-1478

%https://www.mathworks.com/help/robust/ug/robustness-analysis-in-simulink.html

% https://home.cs.colorado.edu/~srirams/papers/robustnessSimulink09.pdf

%https://www.public.asu.edu/~gfaineko/pub/sTaliro_TR.pdf

%https://macsphere.mcmaster.ca/handle/11375/26126

%https://openai.com/blog/adversarial-example-research/

%\textbf{Robustness Analysis.}
%use servey or Justyna'a papepr
System robustness is considered an important engineering principle and has different implications on different development artifacts~\cite{shahrokni2013systematic}. For example, to ensure robustness, system requirements should account for behaviours that allow a system to leave each of its failure states~\cite{roast,jaffe1990software}; system design and implementation should include extensive error-handling~\cite{ding2005dependency, issarny1992exception}; and, the development process should have a mechanism to predict and prevent robustness issues~\cite{ait2003robustness, laranjeiro2010applying}. %Among robustness considerations at the different stages of development , 
Our work falls under the umbrella of robustness testing~\cite{justyna, ballista}, which aims to determine whether a system functions properly in the presence of erroneous inputs or stressful environmental circumstances \cite{shahrokni2013systematic}. A common technique for robustness testing is fault injection. %expected to detect failures in fault-injected systems. 
Fault injection for robustness testing spans both software (e.g., code mutation~\cite{looker2005comparison}) and hardware (e.g., electromagnetic interferences~\cite{hayashi2011non} and power-supply disturbances~\cite{landi2008performances}). For instance, Li et al.~\cite{li1994software} evaluate the robustness of a telecommunication system by injecting software faults into the service manager; and, Barbosa et al. \cite{barbosa2007verification} employ fault injection to evaluate the robustness of third-party components at the interface level. For finding non-robust behaviours, instead of using fault injection, our approach relies on a robustness measure inspired by the search-based testing literature~\cite{harman2009theoretical, liu2017improving}. In addition to detecting individual cases of non-robustness, our approach identifies conditions under which a system \hbox{exhibits non-robust behaviours.}

The closest work to ours is \textsc{S-TaLiRo}~\cite{staliro, montecarlo}, which is a robustness testing tool for cyber-physical systems specified in Simulink. Through globally minimizing a robustness measure, \textsc{S-TaLiRo} generates counterexamples to a Simulink model's temporal-logic requirements. The robustness measure employed by \textsc{S-TaLiRo} is the degree of perturbation that a Simulink model can withstand without changing the truth value of its specifications (expressed in temporal logic) \cite{fainekos2009robustness, taliro, staliro, montecarlo}. In addition, a model satisfies (resp. dissatisfies) a specification \textit{robustly} if its robustness measure is above (resp. below) zero \cite{fainekos2009robustness, staliro, taliro, montecarlo}.
While we adopt the general concept of robustness measure from \textsc{S-TaLiRo}, our work is different
%from \cite{staliro, taliro, montecarlo} and \cite{fainekos2009robustness} 
in three main ways: (1)~%Even though we have adopted the concept of robustness measure from \textsc{S-TaLiRo}, 
Our robustness measure is inspired by fitness computation in the search-based software testing (SBST) literature and differs from the temporal-logic robustness metric  used by \textsc{S-TaLiRo}. 
(2)~%Even though \textsc{S-TaLiRo} introduces the concept of robust versus non-robust behaviours, 
\textsc{S-TaLiRo} focuses exclusively on falsification of Simulink models, i.e., identification of requirement failures for Simulink models. In contrast, our approach is applied to network traffic-shaping systems. (3)~Similar to most testing tools, \textsc{S-TaLiRo} generates individual test cases, whereas our approach is able to find ranges on input variables for non-robust behaviours.

\vspace*{-.1cm}
\section{Lessons Learned}
\label{sec:lesson}
\vspace*{-.05cm}
In this section, we reflect on two lessons learned from our collaboration with RabbitRun and the development of SOHOSim and ENRICH. 

%The lessons concern the benefits of constructing system-level simulators, the nature of non-robustness in our study context, and inspirations for future work based on our current results. We believe our lessons would be most relevant for software engineering researchers and practitioners working in the domains of network systems, cyber-physical systems, and self-adaptive systems.

\textbf{Simulation as a way to discover unknown/undocumented behaviours.}
SOHOSim, in addition to enabling non-robustness analysis, helped our industry partner with the identification of unknown or undocumented behaviours. In particular,  the multiple rounds of experimentation we conducted with RabbitRun using SOHOSim led to the following observations: (1)~The total bandwidth allocated to CAKE should be configured such that it is not limiting the actual maximum network bandwidth. (2)~The class priorities in CAKE are inversely related with the threshold ranges for the classes. (3)~The quality of experience in each CAKE class depends not only on the priority of that class, but also on the bandwidth of the flows passing through the class.  A key reason we could derive such high-level observations, while treating NTSS as an opaque box, is that SOHOSim is a \emph{system-level} simulator. The main lesson learned here is that system-level simulators, in addition to fulfilling their analytical purpose (in our case, analysis of non-robustness), can be useful tools for exploration and identifying unknown/undocumented behaviours of complex systems. 
 
%The multiple rounds of experimentation we conducted with RabbitRun using SOHOSim led to a number of observations related to the behaviours of CAKE~\cite{cakepaper}, i.e., the NTSS implementation that was the subject of our evaluation.

%This enabled the domain experts to conveniently feed SOHOSim with numerous real-world scenarios, in turn providing the opportunity to glean overall trends of behaviour.

%As for the informativeness of our results, RabbitRun expert noted that the input ranges obtained for each NTSS class are understandable to them, and generally consistent with their overall intuition about NTSS. However, they stressed that from the results,  they observed a number of new behaviours that were not known to them before and could not be obtained from the documentation or the public knowledge of CAKE~\cite{cakepaper}, the NTSS algorithm used by RabbitRun. In particular, how NTSS should be configured so that it is not limiting the maximum network bandwidth. Also, the way priorities of NTSS classes interplay with the threshold ranges of each NTSS class.

%When not performing the job of the limiter, results were as expected from the prioritization standpoint, even though we observed the code had X. Being the limiter has adverse effects and may skew results.

\textbf{Non-robustness does not imply faultiness.} For network systems, non-robustness is inevitable when bandwidth is constrained. As demand increases, the available bandwidth is eventually exceeded. No matter how well-designed an NTSS is, if overwhelmed, its quality of service eventually transitions to being robustly bad. This means that non-robustness is to be expected as one crosses the boundary between robustly good and robustly bad regions. An important lesson in our study context is that non-robust regions of the input space are best \emph{not} treated as faulty regions; that is, the existence of non-robust regions should not prompt fixes to the NTSS implementation. Instead, non-robust regions should be treated as situations that applications should attempt to steer clear of. 

Taking the above lesson one level further, we believe that when non-robustness cannot be avoided through predefined static mappings, one needs more advanced safeguards, e.g., dynamic reconfiguration and self-adaptation at run-time, to preserve the quality of experience for as long as theoretically feasible. Existing NTSS are not yet equipped with such dynamic features. This presents interesting opportunities for applying ideas from self-adaptive systems to NTSS. For future work, we plan to investigate the use of our existing non-robustness analysis technique for developing NTSS that can dynamically guide an NTSS out of non-robust regions.

\section{Conclusion}
\label{sec:con}
We proposed an approach that combines software testing and machine learning to generate input constraints that characterize a system's non-robust behaviours. We instantiated and empirically evaluated our approach over a novel case study from the network domain. %, i.e., a network traffic shaping system (NTSS),
This case study, which is concerned with a network traffic-shaping system, was conducted in collaboration with an industry partner, RabbitRun Technologies.  Our approach accurately characterizes non-robust test inputs of NTSS by achieving a precision of 84\% and a recall of 100\%,
significantly outperforming a standard baseline.

\section*{Acknowledgment}
We gratefully acknowledge funding from RabbitRun Technologies Inc., and NSERC of Canada under the Alliance, the Discovery and Discovery Accelerator programs.

\bibliographystyle{IEEEtran}
\balance
\bibliography{paper}

\end{document}